# In-situ tunable giant electrical anisotropy in a grating gated AlGaN/GaN two-dimensional electron gas

Ting-Ting Wang[1,2], Sining Dong[1,2,*], Chong Li[1,2], Wen-Cheng Yue[1,2], Yang-Yang Lyu[1,2], Chen-Guang Wang[1,2], Chang-Kun Zeng[1], Zixiong Yuan[1,2], Wei Zhu[3], Zhi-Li Xiao[4,5], Xiaoli Lu[6], Bin Liu[1], Hai Lu[1], Hua-Bing Wang[1,2,7], Peiheng Wu[1,2,7], Wai-Kwong Kwok[4] and Yong-Lei Wang[1,2,7,*]

[1]*School of Electronic Science and Engineering, Nanjing University, Nanjing, 210023, China*

[2]*Research Institute of Superconductor Electronics, Nanjing University, Nanjing, 210023, China*

[3]*Key Laboratory for Quantum Materials of Zhejiang Province, School of Science, Westlake University, Hangzhou 310024, China*

[4]*Materials Science Division, Argonne National Laboratory, Argonne, Illinois 60439, USA*

[5]*Department of Physics, Northern Illinois University, DeKalb, Illinois 60115, USA*

[6]*School of Microelectronics, State Key Discipline Laboratory of Wide Bandgap Semiconductor Technology, Xidian University, Xi'an 710071, China*

[7]*Purple Mountain Laboratories, Nanjing, 211111, China*

* Correspondence to: sndong@nju.edu.cn; yongleiwang@nju.edu.cn

**Abstract**

**Materials with in-plane electrical anisotropy have great potential for designing artificial synaptic devices. However, natural materials with strong intrinsic in-plane electrical anisotropy are rare. We introduce a simple strategy to produce extremely large electrical anisotropy via grating gating of a semiconductor two-dimensional electron gas (2DEG) of AlGaN/GaN. We show that periodically modulated electric potential in the 2DEG induces in-plane electrical anisotropy, which is significantly enhanced in a magnetic field, leading to an ultra large electrical anisotropy. This is induced by a giant positive magnetoresistance and a giant negative magnetoresistance under two orthogonally oriented in-plane current flows, respectively. This giant electrical anisotropy is in-situ tunable by tailoring both the grating gate voltage and the magnetic field. Our semiconductor device with controllable giant electrical anisotropy will stimulate new device applications, such as multi-terminal memtransistors and bionic synapses.**

**1**. Two dimensional (2D) systems with dramatic in-plane anisotropy are of special interest because they have great potential for realization of multi-terminal memtransistors which could enable complex neuromorphic learning[1-3]. A strong electrical anisotropic behavior is ideal for mimicking the different connection strengths among biological neural network synapses[4-6]. Recently, large efforts have been made on searching material systems with electrical anisotropy, such as in black phosphorus[4,7], GaTe[8] and WSe$_2$[1]. However, the intrinsic electrical anisotropy in these 2D materials is small because of the nearly identical effective mass at K and K′ valleys in different orientations[9]. On the other hand, electronic nematicity or stripe-like phase, which is usually associated with a spontaneous symmetry breaking, leads to discrepancy of conductivity along different directions[10]. For example, in the EuO/KTaO$_3$ (111) superconducting interface, a distinct stripe phase produces large anisotropic transport near the onset of superconductivity[11]. Furthermore, in two-dimensional electron gas (2DEG) systems, the in-plane spatially modulated potential naturally formed during certain growth procedure[12] and the strain-induced grating on the sample surface[13] result in anisotropic behavior, where stripe-like periodic potential/electron population are formed. Therefore, it could be possible to in-situ manipulate the electrical anisotropy of a 2DEG system by artificially introduced electrical potentials, which would be consequential toward neuromorphic device applications.

**2**. Here we explore the in-situ tuning of the electrical anisotropy in a grating gated 2DEG system of AlGaN/GaN heterostructure. Previous studies of the grating gated 2DEG systems, combined with field-effect transistors, were widely used for terahertz

(THz) applications[14-16]. However, the electrical anisotropy properties of the grating gated 2DEG systems are largely overlooked. Herein, we conducted experiments of the electrical current angle dependence of the magnetoresistance (MR) in a grating gated 2DEG system. We achieved an extremely large anisotropic magnetoresistance (AMR) effect, which is continuously tunable by the applied gating voltage and magnetic field. Our findings provide an economical and practical way to design large in-plane electrical anisotropy device.

**3**. The AlGaN/GaN heterostructure films [Fig. 1(a)] used in this work were purchased from Suzhou Enkris Semiconductor Inc. The AlGaN/GaN films with a 14.5 nm capping layer were patterned into a cross bar geometry [Fig. 1(b)] by using photolithography and inductively coupled plasma (ICP) etching. The currents $I$ can be applied in any desired in-plane orientations, which allows rapid and convenient measurement of the in-plane conductance anisotropy[17,18]. The bridge width of the cross bar is $w = 60$ $\mu$m. The 2DEG is at 20 nm below the sample surface. Electrodes of Ti(30 nm)/Al(150 nm)/Ni(50 nm)/Au(100 nm) with Ohmic contacts were made by electron beam evaporating, followed by a high temperature annealing at 850 °C for 30 seconds. At last, grating gates made of 30nm Ti/100nm Au, were fabricated on top of the sample as shown in Fig. 1(a). The width of each grating stripe line is 5 $\mu$m and the period is 10 $\mu$m. The gate stripe lines are alternatively separated in to two groups to produce periodically modulated potential in the 2DEG, as shown in Figs. 1(a) and 1(b). One group of the gating electrodes are grounded, while a gate voltage $V_g$ is applied to the other group of the gating electrodes to regulate the potential amplitude. We define that

$\theta = 0°$ when the current flows perpendicular to the stripe gate lines and $\theta = 90°$ for the parallel configuration [Fig. 1(b)]. The longitudinal and transverse resistances are obtained from the voltage components parallel and perpendicular to the current flow, respectively. More details about our measurement method are provided in the supplementary material. All the measurements were carried out at a current of 100 $\mu$A. The magnetic field for magnetoresistance measurements was applied along the normal direction of the film plane. We conducted conventional resistance measurements in a Cryogen Free Superconducting Magnet Systems (CRYOMAGNETICS, INC.).

**4**. Figure 1(c) presents the polar coordinate profile of the longitudinal resistance as a function of the current orientation $\theta$ under various gate voltages under zero magnetic field and at 1.8 K. When the negative grating gating voltage $V_g$ is applied from 0 V to -4 V, the resistances for all the current directions increase gradually due to the decrease of carrier density. At $V_g = -4$ V, the resistance for $\theta = 0°$ is 52.5 $\Omega$ which is slightly larger than $R = 48.4$ $\Omega$ for $\theta = 90°$. Here we define the electrical anisotropy factor $\sigma = R_{max}/R_{min}$, where $R_{max}$ and $R_{min}$ are the maximum and minimum of the measured resistances, respectively. It can be obtained $\sigma = 1.085$ at $V_g = -4$ V, indicating a very tiny electrical anisotropy induced by grating gating under zero magnetic field. Surprisingly, this uniaxial anisotropy becomes dramatically enhanced when an out-of-plane magnetic field is applied. Figure 1(d) shows $R$ vs. $\theta$ curves at $V_g = -4$ V in several selected magnetic fields. A giant electrical anisotropy is observed at a high magnetic field. The longitudinal resistances for the current directions near $\theta = 0°$ (180°) exhibit quadratic increases with increasing the magnetic field from zero to 9 T. On the contrary, the resistances of $\theta = 90°$ (270°) decrease significantly with the magnetic field. Figure 1(e) shows the angular $\theta$ dependence of longitudinal resistance under various applied gate voltages and at a magnetic field of 9 T. One can see that the gating effect can tremendously influence the anisotropy of the electrical properties for this system under a high out-of-plane magnetic field. Thus, the large electrical anisotropy is in-situ tunable by both the grating gate voltage and the out-of-plane magnetic field.

5.  The resistances change with the magnetic and electrical fields in Fig. 1(d) and 1(e), respectively, except for several certain current directions (e.g. $\theta \sim 75°, 105°, 255°, 285°$), which separate the positive magnetoresistance (PMR) region and the negative magnetoresistance (NMR) region. To directly demonstrate the magnetoresistance effects, we measured the field dependent magnetoresistance at various current orientations [Fig. 2(a)]. The magnetoresistance is defined as MR = $[R(B) - R(0)]/R(0)$ × 100%, where $R(B)$ and $R(0)$ are the resistance in a magnetic field $B$ and that in a zero field, respectively. The MR curve shows a nearly zero magnetoresistance for $\theta = 75°$ at ±9 Tesla. The MR for $\theta = 0°$ has a largest PMR without any sign of saturation at magnetic fields as high as 9 T, while the MR for $\theta = 90°$ has a large NMR up to -82% at 9 T. The MRs of our grating gated 2DEG sample present positive and negative extremes at the two orthogonal directions, respectively, leading to a huge electrical anisotropy factor $\sigma$ at high magnetic fields. Figure 2(b) shows the magnetic field dependence of the electrical anisotropy factor at $V_g = -4$ V and $T = 1.8$ K. It appears that $\sigma$ increases exponentially with the magnetic field, and it reaches a value of 125.3 at $B$ = 9 T and is not saturate. When we reduce the grating gate voltage $V_g$ from -4 V to -1

V, as illustrated in Fig. 2(c), it shows a monotonic decrease in $\sigma$, whose value finally becomes to be around 1 at $V_g = 0$ V or further applying a positive $V_g$. Hence, these results demonstrate again that the periodic gating potential formed in the 2DEG channel leads to a dramatic and tunable electrical anisotropy in magnetic fields.

**6**.  To investigate the thermal evolution of the electrical anisotropy of the system, we carried out magnetoresistance measurements at a series of selected temperatures. Figure 3(a) shows the MR vs. *B* curves at various temperatures for $V_g = 0$ V. The MR curve at 1.8 K shows a negative parabolic magnetoresistance as a function of the magnetic field. It reveals Shubnikov-de Haas (SdH) quantum oscillations[19] which is conspicuous for $B > 4$ T. As the temperature increases, the magnitude of the NMR in a given magnetic field gradually decreases and becomes very small at temperatures above 180 K. The MR vs. *B* curves measured under $V_g = -4$ V and $\theta = 0°$, as shown in Fig. 3(b), are dramatically different from that without a grating gating voltage. Significant PMR effects are obtained for all the temperatures we measured, although the magnitude also decreases with the temperature. Fig. 3(c) presents the MR curves measured under $V_g = -4$ V and $\theta = 90°$ at various temperatures. One can observe huge NMR effect at the low temperatures. At high temperatures, e.g. at 250 K, the MR effect is still remarkable with a value of -27.6% at $B = \pm 9$ T. As a comparison, the MR is only -12% at the lowest temperature 1.8 K without any gating voltage applied. Figures. 3(b) and 3(c) show that the SdH oscillations are suppressed by applying the periodic gate voltages for both current orientations. The temperature dependence of electrical anisotropy factor $\sigma$ at $V_g = -4$ V and $B = 9$ T is shown in Fig. 3(d). The factor $\sigma$ undergoes an exponential decay with increasing temperatures. At 250 K $\sigma = 4$ is still a remarkable anisotropic coefficient as compared to a natural material.

7. Previous investigation shows that the same 2DEG sample covered by a uniform gate pad displays a PMR of 10% under a flat gating voltage of -4 V[20]. Therefore, the giant PMR and NMR observed in the sample with the grating gate are ascribed to the artificial periodic potential in the 2DEG channel, which has notable difference on the transport properties for the currents with different directions. For $\theta = 0°$ (the current is perpendicular to the periodic gate stripes) the PMR is approximately quadratic as a function of magnetic field. This behavior is similar as that found in GaAs/AlGaAs 2DEG systems with a periodic potential at the low magnetic field range[21,22], which can be explained by considering the classical trajectories of electrons in a periodic potential[21]. The suppression of the SdH oscillations was also observed in the GaAs/AlGaAs 2DEG system, which results from the broadening of the Landau levels by the periodic potential[21]. The PMR for $\theta = 0°$ in the AlGaN/GaN 2DEG system probably originates from a similar mechanism. However, differing from the quadratic PMR effect in the GaAs/AlGaAs, which only exists in the weak magnetic field (below a few hundred milli-Tesla[22]), the quadratic PMR phenomenon is non-saturating up to 9 T in our AlGaN/GaN sample. A non-saturating extremely large PMR in the multiband topological quantum materials, such as in a Weyl semimetal $WTe_2$ crystal, was attributed to the balanced electron–hole populations[23], which is clearly not the case in our system. Furthermore, the periodicity of the 1D grating potential in this work is 20 $\mu$m which much larger than that of sub micrometers in the GaAs 2DEG[21,22]. Aside from that, the electron mobility of GaAs/AlGaAs 2DEG is over $10^6$ cm$^2$ V$^{-1}$ s$^{-1}$ which is 100 times larger than that of the AlGaN/GaN system used in this work (see Fig. 3(d)).

**8**.  The uniform AlGaN/GaN 2DEG without a grating gating shows a notable NMR effect, which originates from the memory effect induced by interface roughness scattering[20]. This NMR effect is suppressed by applying a uniform gating voltage[20]. In contrast, the grating gate voltage significantly enhances the NMR effect for $\theta = 90°$ in this work [Figs. 3(a) and 3(c)]. The MR vs. *B* curves show a steep drop at low fields [Fig. 3(c)], which is similar to the colossal NMR discovered in a GaAs/AlGaAs quantum well[24]. The giant NMR in the GaAs quantum well was attributed to a hydrodynamic mechanism that magnetic field decreases the diagonal viscosity in ultrahigh-mobility 2D electrons[25]. Large NMR was also observed in the width limited channels of ultrapure 2D metal PdCoO$_2$[26] and high mobility graphene[27] based on electron viscous flow. Electrical gating alters both the carrier density and the mobility in AlGaN/GaN 2DEG system[20]. Hence, the grating gate produces stripe channels alternately containing two different carriers. However, the mean free path of electrons in this AlGaN/GaN 2DEG system is about 0.4 $\mu$m[20] that is much smaller than the period 20 $\mu$m of the grating gate, which is incompatible with the requirement of electron viscous flow in the high purity 2D samples[25]. Nevertheless, the PMR at $\theta = 0°$ and the NMR at $\theta = 90°$ in our grating gated 2DEG sample lead to incredible in-plane uniaxial electrical anisotropy in a high magnetic field. Future investigations, such as the period dependence and grating stripe width dependence of the magnetoresistance, are still needed for understanding the detailed microscopic mechanism of the giant PMR and NMR effects observed in this work.

**9**. To further demonstrate the in-plane anisotropy effect, we present in Fig. 4(a) the angle dependence of the transverse (or Hall) resistance $R_T$ obtained simultaneously with the measurement of the longitudinal resistance shown in Fig. 1(d). As negative gating voltages are applied, clear uniaxial anisotropic transverse resistance is observed. The current direction of the minimum transverse resistance has a 45° shift compared with that for longitudinal resistance. This shift is robust and is independent with the magnetic fields as shown in Fig. 4(b). At zero applied magnetic field this two-fold anisotropy at 45° induces sign reversal in the orientation dependent transverse resistance (see the zoom-in data for $B = 0$ T in Fig. 4(c)). Such a behavior of the transverse resistance was reported in the copper oxide superconductor $La_{2-x}Sr_xCuO_4$ film with an electronic nematicity[28]. Therefore, our 2DEG with grating gating could mimic a tunable material with an artificial electronic nematic phase. Beyond the result shown in the naturally existing nematicity in the copper oxide superconductor[28], our results demonstrate that the magnetic field significantly enhances the electrical anisotropy in our artificial nematicity of the grating gated AlGaN/GaN 2DEG. These electrical anisotropy properties can be directly revealed by our conductance tensor calculations, in which the normalized longitudinal and transverse resistances are in good agreement with our experiment results (see supplementary material Fig. S2). Our experiments and calculations indicate the anisotropy of the transverse resistance is an intrinsic and universal phenomenon in an electrical anisotropic system, which could benefit applications, such as artificial biological neural network synapse devices[1,4-6].

10. In conclusion, the grating gate induces a periodic potential in the 2DEG of AlGaN/GaN, which creates in-situ tunable artificial electronic nematicity with tailored electrical anisotropy. Extremely large PMR and NMR behaviors have been obtained when current is rotated along different in-plane directions in a cross-bar sample. This results in a dramatically large in-plane electrical anisotropy as revealed by both the longitudinal resistances and the transverse resistances. Our findings could be advantageous for designing programmable devices, such as multi-terminal memtransistors, for complex neuromorphic applications. Room temperature high mobility materials, such as graphene, may also be a good candidate, whose electrical anisotropy can be artificially manipulated at much higher temperatures.

See the supplementary material for details of the measurement method and the conductance tensor calculations.

**AUTHOR DECLARATIONS**

**Conflict of Interest**

The authors have no conflicts to disclose.

This work is supported by the National Key R&D Program of China (2018YFA0209002 and 2021YFA0718802), the National Natural Science Foundation of China (61971464, 61727805 and 11961141002), Jiangsu Excellent Young Scholar program (BK20200008), Jiangsu Shuangchuang program, the Fundamental Research

Funds for the Central Universities and Jiangsu Key Laboratory of Advanced Techniques for Manipulating Electromagnetic Waves. The analysis of magnetoresistance and manuscript editing (Z.-L.X. and W.-K.K.) was supported by the U.S. Department of Energy, Office of Science, Basic Energy Sciences, Materials Sciences, and Engineering. Z.-L.X. also acknowledges support from the National Science Foundation under Grant No. DMR-1901843 for his efforts on data analysis.

**DATA AVAILABILITY**

The data that support the findings of this study are available from the corresponding author upon reasonable request.

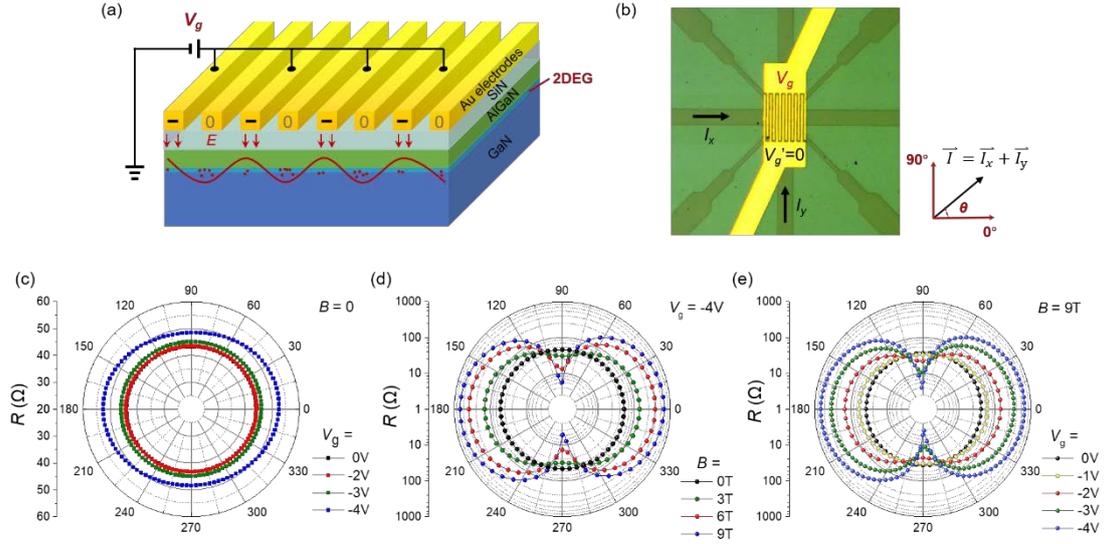

**Fig. 1** (**a**). Schematic diagram of the 2DEG sample with grating metal gate electrodes. The meandering red curve illustrates periodical modulated potential, which can be regulated through $V_g$. (**b**). Optical image of the sample. The current orientation $\theta$. (**c**)-(**e**). Angular $\theta$ dependence of longitudinal resistance at various gate voltages in a zero magnetic field (c), in various magnetic fields for $V_g$ = -4 V (d) and under various grating gate voltages in a magnetic field of $B$ = 9 T (e). These results were measured at a temperature of $T$ = 1.8 K.

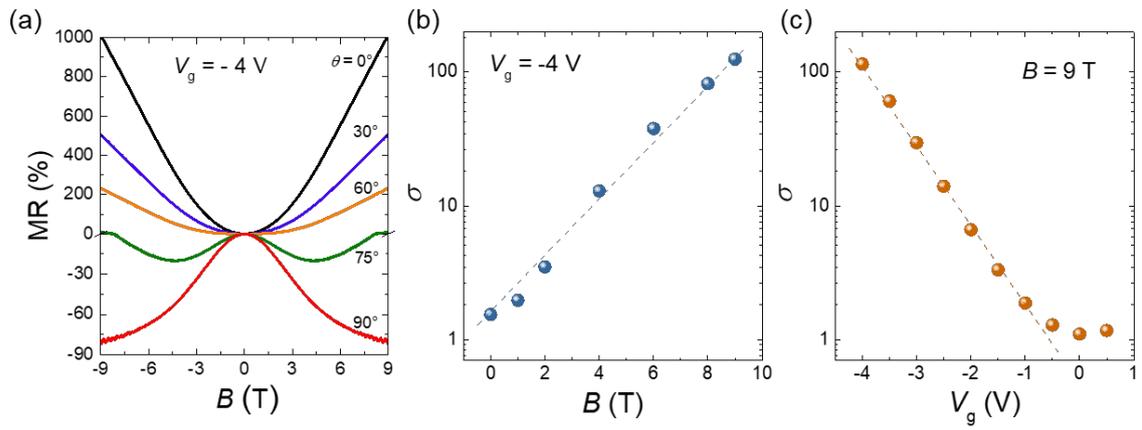

**Fig. 2** (**a**). Magnetoresistance as a function of magnetic field for various current directions at $V_g$ = -4 V and $T$ = 1.8 K. (**b**) and (**c**). Electrical anisotropy factor $\sigma$ as a function of magnetic field (b) and grating gating voltage (c), respectively. The dash lines are guides for eyes.

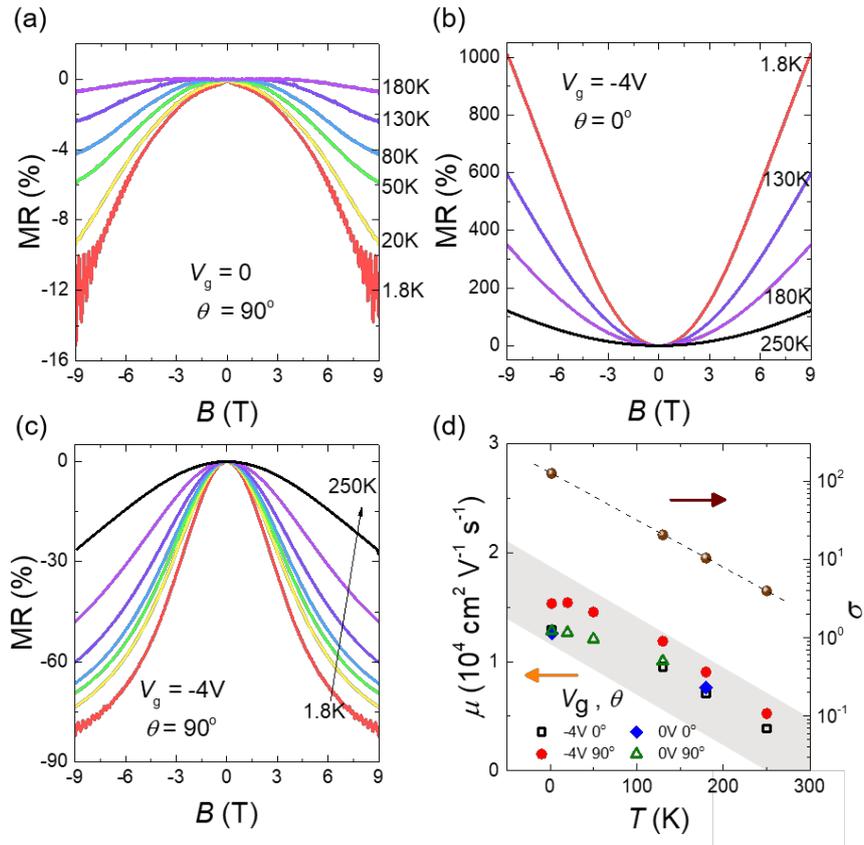

**Fig. 3 (a)**. Magnetoresistance as a function of magnetic field at various temperatures with $V_g = 0$ V. **(b)** and **(c)**. Magnetoresistance as a function of magnetic field at various temperatures with $V_g = -4$ V for $\theta = 0°$ (b) and $\theta = 90°$ (c), respectively. **(d)**. Mobility $\mu$ as a function of temperature for various gating voltages and current directions (lower left panel), and semi-logarithmic plot of temperature dependent electrical anisotropy factor $\sigma$ at $V_g = -4$ V and $B = 9$ T (upper right panel, the dash line is guide for eyes).

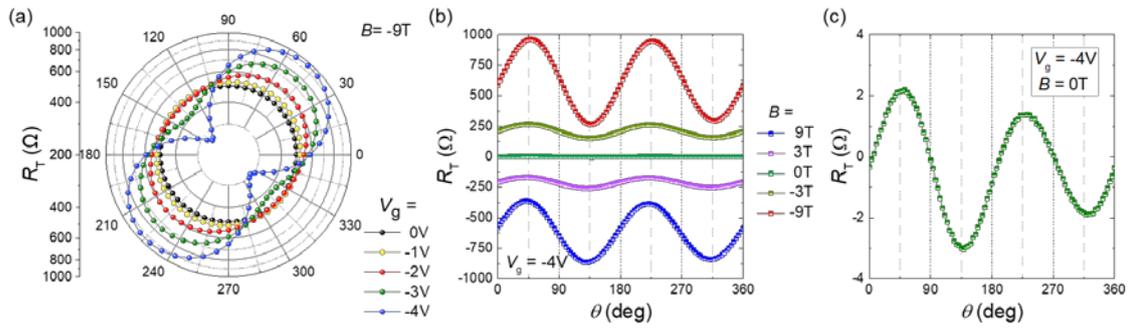

**Fig. 4 (a)** and **(b)**. Angular $\theta$ dependence of the transverse resistance at various grating gate voltages in $B = -9$ T (a) and in different magnetic fields for $V_g = -4$ V (b). **(c)**. Enlarge view of $R_T$ vs. $\theta$ curve at $B = 0$ in (b). The results were measured at $T = 1.8$ K.